\documentclass[aps,prd,preprint,superscriptaddress,tightenlines,nofootinbib]{revtex4}

\usepackage{graphicx}
\usepackage{dcolumn}
\usepackage{bm}

\begin{document}

\preprint{CLEO CONF 06-3}   

\title{Study of Particle Production in Quark vs.
Gluon Fragmentation at $\sqrt{s}\sim$10 GeV}

\thanks{Submitted to the 33$^{\rm rd}$ International Conference on High Energy
Physics, July 26 - August 2, 2006, Moscow}

\author{R.~A.~Briere}
\author{I.~Brock~\altaffiliation{Current address: Universit\"at Bonn; Nussallee 12; D-53115 Bonn}}
\author{J.~Chen}
\author{T.~Ferguson}
\author{G.~Tatishvili}
\author{H.~Vogel}
\author{M.~E.~Watkins}
\affiliation{Carnegie Mellon University, Pittsburgh, Pennsylvania 15213}
\author{J.~L.~Rosner}
\affiliation{Enrico Fermi Institute, University of
Chicago, Chicago, Illinois 60637}
% ...
\collaboration{CLEO Collaboration} 
\noaffiliation

\begin{abstract}
Using data collected with the CLEO~III detector at the Cornell Electron Storage
Ring, 
we have compared proton, lambda (charged conjugate
modes are implicit) and meson
($\phi$ and $f_2(1270)$)
production observed in gluon fragmentation vs. quark fragmentation.
Two studies have been conducted: in the first, we corroborrate
previous per-event particle
yields in $\Upsilon\to ggg$
vs. $e^+e^-\to q{\overline q}$. In the second, we compare
particle production in the photon-tagged process
$\Upsilon({\rm 1S}) \rightarrow gg\gamma$ with that
in $e^+e^- \rightarrow q{\bar q}\gamma$ events. 
For each particle,
we determine the `enhancement' ratio, defined
as the ratio of
particle yields per gluon
fragmentation event compared
to quark fragmentation event. Thus
defined, an enhancement of 1.0 implies
equal per-event production in both gluon and quark 
fragmentation.
In the photon-tagged analysis,
we find an enhancement of
order unity for protons and approximately 1.5 for
$\Lambda$'s.
This measured proton enhancement rate is supported by a study
of baryon production in $\chi_{b,2}\to p+X$ relative
to $\chi_{b,1}\to p+X$. 
The production of
mesons having masses of order 1 GeV ($\phi$ and $f_2(1270)$) are
found to be also approximately the same in $gg\gamma$ vs. 
$q{\bar q}\gamma$ fragmentation.
Overall, per-event
baryon production in two-gluon fragmentation is considerably
smaller than that observed in
three-gluon decays of the $\Upsilon$(1S). 
Our results for baryon
production are inconsistent with the predictions of the JETSET (7.4)
fragmentation model.
The results presented in this document are preliminary.
\end{abstract}

\newpage

\maketitle

\date{July 24, 2006}
\tighten

{
 \renewcommand{\thefootnote}
 {\fnsymbol{footnote}}
 \setcounter{footnote}{0}
}

\newpage

\setcounter{footnote}{0}

\section{Introduction}
Understanding
hadronization, or the process by which elementary partons (gluons and quarks)
evolve into mesons and baryons, 
is complicated by its intrinsically
non-perturbative nature. 
Due to the fact that gluons carry two color indices whereas
quarks carry only one, the intrinsic
gluon-gluon coupling strength ($C_A$=3) is larger than the
intrinsic quark-gluon coupling strength ($C_F$=4/3).
Radiation of secondary
and tertiary gluons is therefore expected to be
more likely when hadronization is initiated by
a gluon rather than by a quark. This results in a greater number of 
final state hadrons as well as a larger average transverse momentum
in the former case compared to the latter case. In the limit
$Q^2\to\infty$, the ratio of the number of hadrons produced in
gluon-initiated jets to the number of
hadrons produced in quark-initiated jets is expected, in
lowest order, to approach the
well-known color-counting ratio $C_A/C_F$=9/4\cite{r:theory76}.

Many experiments have searched for, and found,
multiplicity and jet shape differences between quark and
gluon fragmentation.
At $Z^0$ energies, {\it q\=qg} events
are distinguished by their three-jet topology. Within such events,
quark and
gluon jets can be separated by a variety of techniques including vertex 
tagging.  Because 
gluons rarely fragment into heavy quarks, they will produce jets that 
form a vertex
at the $e^+e^-$ interaction
point. Quark jets, to the contrary, tend to form a detached
vertex when the jet contains
a long-lived bottom or charm quark. For light-quark events with
gluon radiation, however,
the assignment of final state hadrons to the initial
state partons is generally more ambiguous and often relies on Monte Carlo simulations
to determine the fraction of times that an observed hadron is correctly
traced to a primary parton.

Within the limits of their precision, previous studies at
SLD found inclusive production of pions, kaons and protons to
be equivalent for gluon-tagged and quark-tagged jets\cite{r:SLD01}.
OPAL has measured charm production at the level
of $(3.20\pm0.21\pm0.38)$\% in gluon jets\cite{r:OPALincl04,r:OPALcharm99},
more than an order of magnitude smaller than the rate observed
in quark jets at the $Z^0$. ALEPH\cite{r:ALEPHb99} and DELPHI\cite{r:DELPHIb99}
both measured inclusive bottom production in gluon-tagged jets
to be $2-3\times 10^{-3}$, again considerably smaller than
that expected from charge counting in quark fragmentation.
Most directly comparable to our current work, OPAL has
also compared inclusive $K^0_{\rm s}$ and $\Lambda$ production
in gluon- vs. quark-tagged jets, finding inclusive production
ratios again consistent with unity ($0.94\pm0.07\pm0.07$ and
$1.18\pm0.01\pm0.17$, respectively)\cite{r:OPAL98K0slam}.

The 10 GeV center of mass energy range offers a unique 
opportunity to
probe quark and gluon fragmentation effects, without relying on Monte
Carlo simulation to associate the final state hadrons with an initial state 
parton. The decay 
$\Upsilon(1S) \rightarrow gg\gamma$ allows one to compare the $gg$ system 
in a $gg\gamma$ event with the $q{\bar q}$ system in  
$e^+e^- \rightarrow q{\bar q}\gamma$ events. In both cases,
the system recoiling against the photon consists (to lowest order) of hadrons
that have evolved from either a two-gluon or a quark-antiquark system. 
The properties of the recoil systems can then be compared 
directly.\footnote{Although
there may be gluon radiation from the initial partons, we 
do not distinguish such radiation explicitly in this analysis.
Thus, the states that we are comparing are, strictly speaking,
$gg\gamma$ and $q{\bar q}\gamma$ to lowest-order only; 
additional gluon radiation, to which we are not experimentally sensitive,
may be present in many of the events in our sample.
Without the
ability to adequately identify additional gluons, such
higher-order radiative
effects are therefore implicitly absorbed into the experimental 
measurement.}
Additionally, the radiative transitions from the radially excited
$\Upsilon$ states to the orbitally excited $\chi_b$ triplet offers
an opportunity to further probe fragmentation differences between
decays of the J=0 and J=2 $\chi_b$ states (decaying predominantly
to two gluons) vs. decays of the J=1 state (decaying primarily to
q\=q(g) final states, with the gluon nearly on-shell). Statistical
correlations
between transition photons with
inclusive production of particular final-state
particles (X) allows a measurement of the relative yields of
$gg\to X$:q\=q($g$)$\to$ X.

In this analysis, we focus on the relative production rates of
baryons (protons and $\Lambda$'s) 
and heavy mesons ($\phi$ and $f_2(1270))$ in gluon vs.
quark fragmentation (charge conjugation is implied). 
A previous study noted apparent enhancements in
the production of $\phi$, $\Lambda$ and protons in three-gluon 
decays of the $\Upsilon$(1S)\cite{r:cleo84}, albeit with low
statistics and at no more 
than 2-3$\sigma$ significance. That initial study also
found approximately one unit larger 
charged multiplicity for three-gluon
fragmentation of the $\Upsilon$(1S) compared to q\=q fragmentation at
a comparable center-of-mass energy. With the limited statistics at that
time, the additional unit of multiplicity could entirely be accounted for
by enhanced three-gluonic production of baryons.
We now have sufficient
statistics to re-measure the three-gluon particle production rates,
and also to compare, for the first time, 
inclusive production in two-gluon fragmentation vs.
inclusive production in three-gluon fragmentation. 

Since then, other experimental data on quark/gluon fragmentation differences
in the $\sqrt{s}\sim$10 GeV energy regime
have become available, including (per event):
\begin{enumerate}
\item A suppression of production of open
charm in gluonic decays of the $\Upsilon$ resonances:
$(\Upsilon$(1S)$\to ggg\to D+X)/(e^+e^-\to D+X)\sim$0.02\cite{CLEO-charm-06}. 
\item The observation that fragmentation of 
the J=1 state of the $\chi_b$ triplet 
(decaying to q\=qg) results in 
charm production comparable to the underlying continuum;
no such charm production is observed in
the two-gluon decays of the J=0 or J=2 states\cite{GochaTatishvili}.
\item An enhancement in production of hidden charm in gluonic decays of
the $\Upsilon$ resonances:
$(\Upsilon$(1S)$\to ggg\to J/\psi+X)/(e^+e^-\to J/\psi+X)>$5.3\cite{CLEOpsi04}.
\item Production of deuterons from resonant 3-gluon 
decays of both the 
$\Upsilon$(1S) and $\Upsilon$(2S) at the level of $10^{-3}$; 
no corresponding
production of deuterons is observed from the continuum\cite{CLEOdeuteron06}.
Per event enhancements are $\ge$10.
\item Approximately
equal production of $\eta'$ in gluonic decays of
the $\Upsilon$ resonance compared to 
$\Upsilon$ decays via q\=q:
$(\Upsilon$(1S)$\to ggg\to \eta'+X)/(\Upsilon\to {q\overline q}\to \eta'+X)\sim$2/3,
integrated over momentum\cite{CLEOetaprime02}.
\end{enumerate}

\section{Detector and Data Sample}
The CLEO~III detector\cite{r:CLEOIIIa,r:CLEOIIIb} 
is a general purpose solenoidal magnet spectrometer and
calorimeter. 
This system is very efficient ($\epsilon\ge$98\%) 
for detecting tracks that have transverse momenta ($p_T$)
relative to the
beam axis greater than 200 MeV/c, and that are contained within the good
fiducial volume of the drift chamber ($|cos\theta|<$0.94, with $\theta$
defined as the polar angle relative to the beam axis). Below this 
threshold, the charged particle detection efficiency in the fiducial
volume decreases to 
approximately 90\% at $p_T\sim$100 MeV/c. For $p_T<$100 MeV/c, the efficiency
decreases roughly linearly to zero at a threshold of $p_T\approx$30 MeV/c.
Just within the solenoidal magnet coil is the electromagnetic calorimeter,
consisting of 7800 thallium doped CsI crystals.  The central ``barrel'' region
of the calorimeter covers about 75\% of the solid angle and has an energy
resolution of
\begin{equation}
\frac{ \sigma_{\rm E}}{E}(\%) = \frac{0.6}{E^{0.73}} + 1.14 - 0.01E,
                                \label{eq:resolution1}
\end{equation}
where
$E$ is the shower energy in GeV. This parameterization translates to an
energy resolution of about 2\% at 2 GeV and 1.2\% at 5 GeV. Two end-cap
regions of the crystal calorimeter extend solid angle coverage to about 95\%
of $4\pi$, although energy resolution is not as good as that of the
barrel region. 
The tracking system, RICH particle identification system and calorimeter
are all contained 
within the 1.5 Tesla superconducting coil. 
Flux return and tracking
chambers used for muon detection are located immediately outside the coil and 
in the two end-cap regions.

We use the CLEO-III
data collected at the narrow $\Upsilon$
resonances as a source of 
$gg\gamma$ events and data taken just
below the resonances, as well as the below-4S continuum
($\sqrt{s}$=10.55 GeV)
as a source of
$q{\bar q\gamma}$ events. The $\gamma$ in our
$q{\bar q\gamma}$ sample results primarily from
initial state radiation (ISR)\cite{r:bkqed}.
We compare events for which the fractional photon
energy $x_\gamma=E_\gamma/E_{beam}$ are the same.\footnote{This
deviates from the previous convention, for which the scaling variable
was the
invariant mass of the $gg$ and q\=q 
systems recoiling against the hard photon 
($M_{recoil}$, defined by $M_{recoil}$ =
$\sqrt{4E_{\rm beam}^{2}(1-E_{\gamma}/E_{\rm beam})}$).
We take the difference between results using these two conventions
as a systematic error.}

\subsection{Data Samples Used and Event Selection}
We use data taken at the 
$\Upsilon$(1S), 
$\Upsilon$(2S), 
$\Upsilon$(3S), and 
$\Upsilon$(4S) resonances.

We impose event selection requirements identical to those used
in our previous study of inclusive direct photon production in 
$\Upsilon$ decays\cite{r:shawn}.
Those cuts are designed primarily to supress backgrounds such as two-photon collisions,
QED events (including tau pairs), and beam-gas and beam-wall collisions.  

Luminosity,
event count, and photon yields ($x_\gamma>$0.5) are
given in Table \ref{tab:datasets}.
\begin{table}[htpb]
\caption{\label{tab:datasets}
Summary of data used in analysis. 
For each data set, we track the number of photons per unit
luminosity, as well as the total number of observed 
hadronic events per unit luminosity.  
HadEvts denotes the total 
number of events in each sample identified as hadronic by our 
event selection requirements. The number of photons
having scaled momentum (relative to $E_{beam}$) greater than 0.5
is presented in the last column.}
\begin{tabular}{c|c|c|c|c|c} 
DataSet & Type & Resonance & ${\cal L}$ (${\rm pb}^{-1}$) & HadEvts &
$N_\gamma(x>0.5)$\\ 
\hline 
1S & Data & $\Upsilon$(1S) & 1220  & 22780000 & 219000 \\
2S & Data & $\Upsilon$(2S) & 1070  & 9450000 & 88800 \\
3S & Data & $\Upsilon$(3S) & 1420  & 8890000 & 79500 \\
4S & Data & $\Upsilon$(3S) & 5520  & 18970000 & 165000 \\
1S-CO & Data & $<\Upsilon$(1S) & 144  & 515000 & 5700 \\
2S-CO & Data & $<\Upsilon$(2S) & 312  & 932000 & 10300 \\
3S-CO & Data & $<\Upsilon$(3S) & 185  & 532000 & 5900 \\
4S-CO & Data & $<\Upsilon$(4S) & 2100  & 5680000 & 64700 \\ 
1S & JETSET MC & $\Upsilon$(1S) &  & 1160000 & 9900 \\
2S & JETSET MC & $\Upsilon$(2S) &  & 9190000 & 70000 \\
3S & JETSET MC & $\Upsilon$(3S) &  & 3890000 & 27000 \\
4S & B\=B MC & $\Upsilon$(4S) &  & 8350000 & 300 \\
1S-CO & JETSET MC & $<\Upsilon$(1S) &  & 8170000 & 68100 \\
2S-CO & JETSET MC & $<\Upsilon$(2S) &  & 7610000 & 66600 \\
3S-CO & JETSET MC & $<\Upsilon$(3S) &  & 12850000 & 115000 \\
4S-CO & JETSET MC & $<\Upsilon$(4S) &  & 63630000 & 568000 \\ 
\hline
\end{tabular} 
\end{table}

\subsection{Event Backgrounds}
To determine the characteristics of resonant
$gg\gamma$ events, we must subtract the background arising from non-resonant
$q{\bar q\gamma}$ and 
$e^+e^-\to\tau\tau\gamma$ events produced in continuum
$e^+e^-$ annihilations at
$\sqrt{s}=M_{\Upsilon({\rm 1S})}$. 
This is done by direct scaling of the event samples collected off-
resonance on the nearby continuum.

\begin{figure*}
\flushleft{\includegraphics[width=3.2in,height=3.2in]{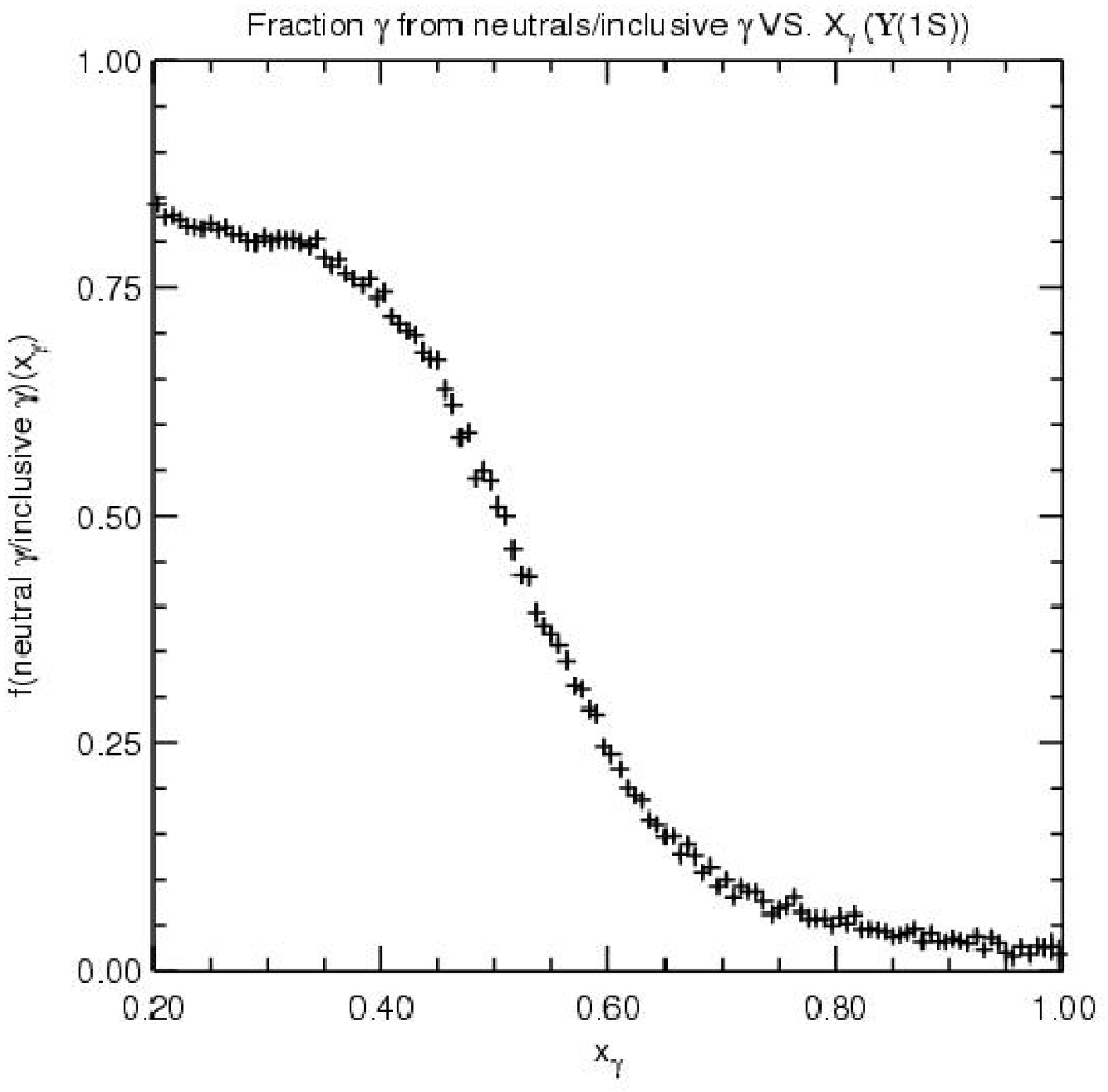}}
\vspace{-3.35in}
\flushright{\includegraphics[width=3.2in,height=3.2in]{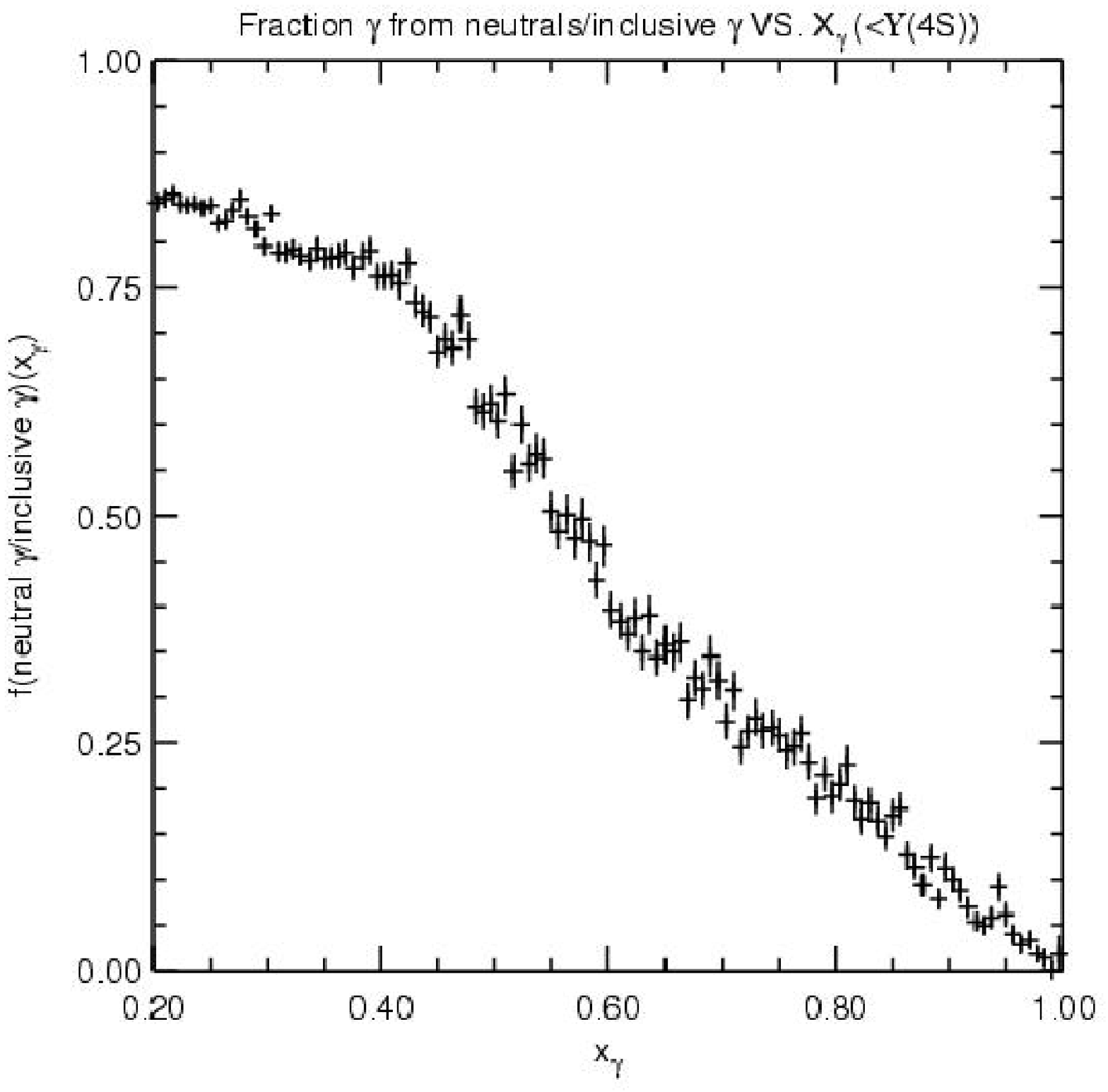}}
     \caption{\label{fig:pi01s} Monte
Carlo prediction for fraction of photons produced
by neutral decays (such as $\pi^0$, $\eta$, $\eta$', and $\omega$)
on the 1S resonance (left) and on the continuum below
the 4S resonance (right). }
\end{figure*}

In order to isolate continuum
$q{\bar q\gamma}$ events, 
$\tau\tau\gamma$ contamination must be explicitly subtracted,
using a Monte Carlo simulation of tau pair events. 
We find that  
$\tau\tau\gamma$ events comprise about 5\% of the $q{\bar q\gamma}$
data sample passing the event selection cuts\cite{r:shawn}.
Beam 
gas and two photon backgrounds were investigated and found 
to be negligibly small.
The photon-tagged sample can also be contaminated by cases where
the high-energy photon is actually a $\pi^0$ daughter.
Figure~\ref{fig:pi01s} illustrates the fraction of photons 
for the on-1S resonance and below-4S continuum, respectively, that are
produced from neutral decays (including not only $\pi^0$ decays but also
$\eta$, $\eta$', and $\omega$ decays) as determined
from Monte Carlo simulations.

Integrated over all tag photon momenta, $\pi^0$ contamination comprises a
$\sim$15\% background to the direct-photon sample.
Monte Carlo simulations also indicate that the $\pi^0$ 
contamination tends to cancel when we take ratios of resonant photon
production to continuum photon production.

\subsubsection{Particle Identification}
Our photon and particle identification procedures are 
identical to those developed in \cite{r:shawn}.
Photon candidates are selected from showers 
with widths and patterns of energy deposition consistent with that of a 
photon, as opposed 
to neutral hadrons
(e.g. merged $\pi^0$'s, $K^0_{\rm L}$, neutrons, etc.). 
To ensure that the 
events are well-contained within the CLEO detector,
we require 
$|\cos\theta_\gamma|<0.707$ ($\theta_\gamma$ defined as before
as the polar angle between the beam axis and the direct photon).
For protons (and antiprotons), we require that charged tracks have specific
ionizatation and also RICH information
consistent with that expected for true protons. For momenta less
than 1 GeV/c, we require that the track particle identification
information be
inconsistent (at the level of $2\sigma$) with that expected for true
pions.
We also require that the proton momentum be greater than 400 MeV, so
as to suppress beam-wall
and fake backgrounds.
For reconstruction of 
$\phi$ (and $f_2$(1270)) from kaons (and pions) we require that 
pairs of opposite charge have particle identification
information consistent with their assumed identities.
For the case of the $f_2$, we
require the pion energy to be greater than 500 MeV, to enhance the 
signal to noise ratio.
Lambdas are identified using the standard
CLEO algorithms for reconstruction of detached vertices.

\subsubsection{Backgrounds to the Proton Sample}
\begin{figure*}
\flushleft{\includegraphics[width=3.5in,height=3.5in]{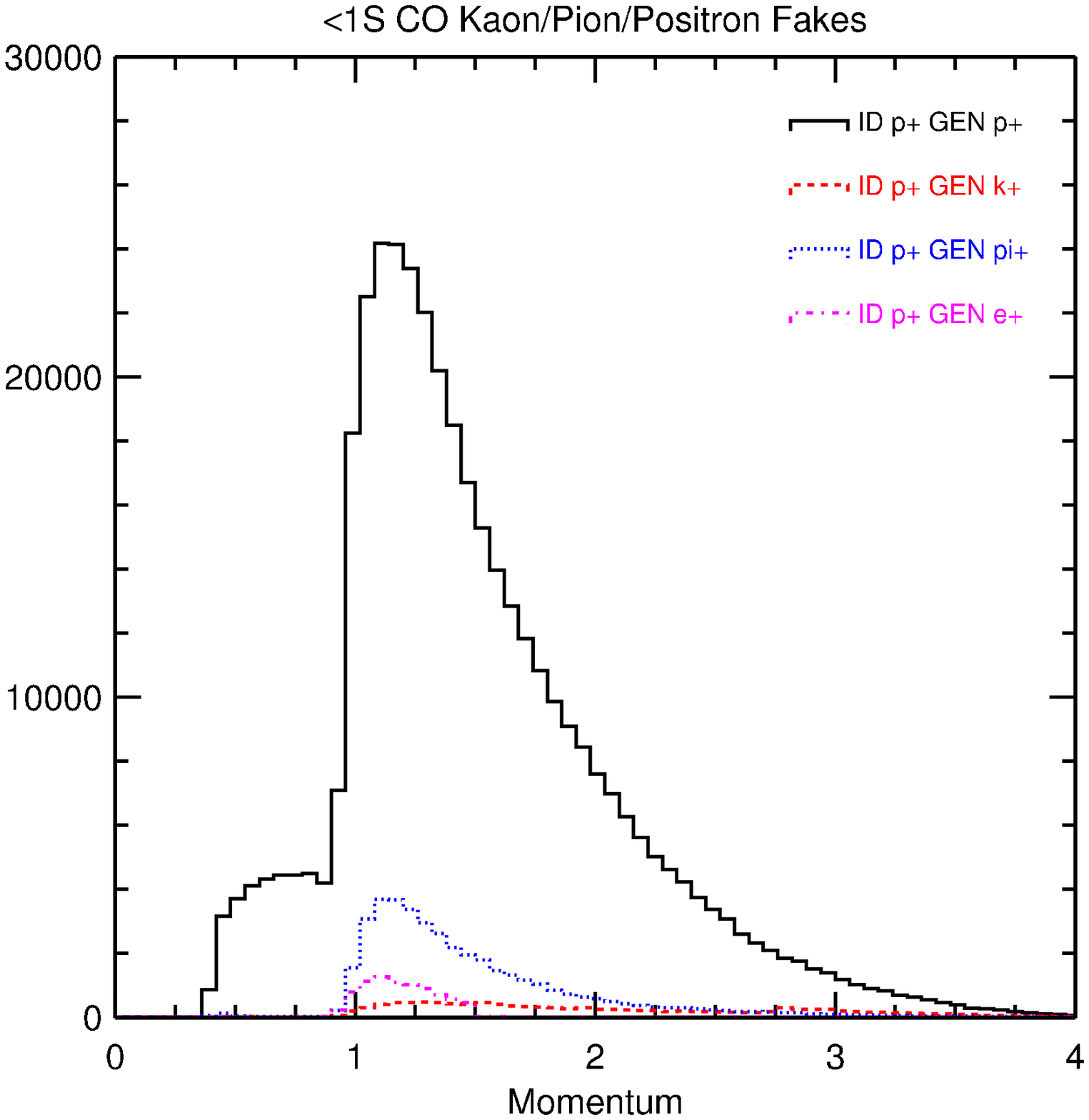}}
\vspace{-3.65in}
\flushright{\includegraphics[width=3.5in,height=3.5in]{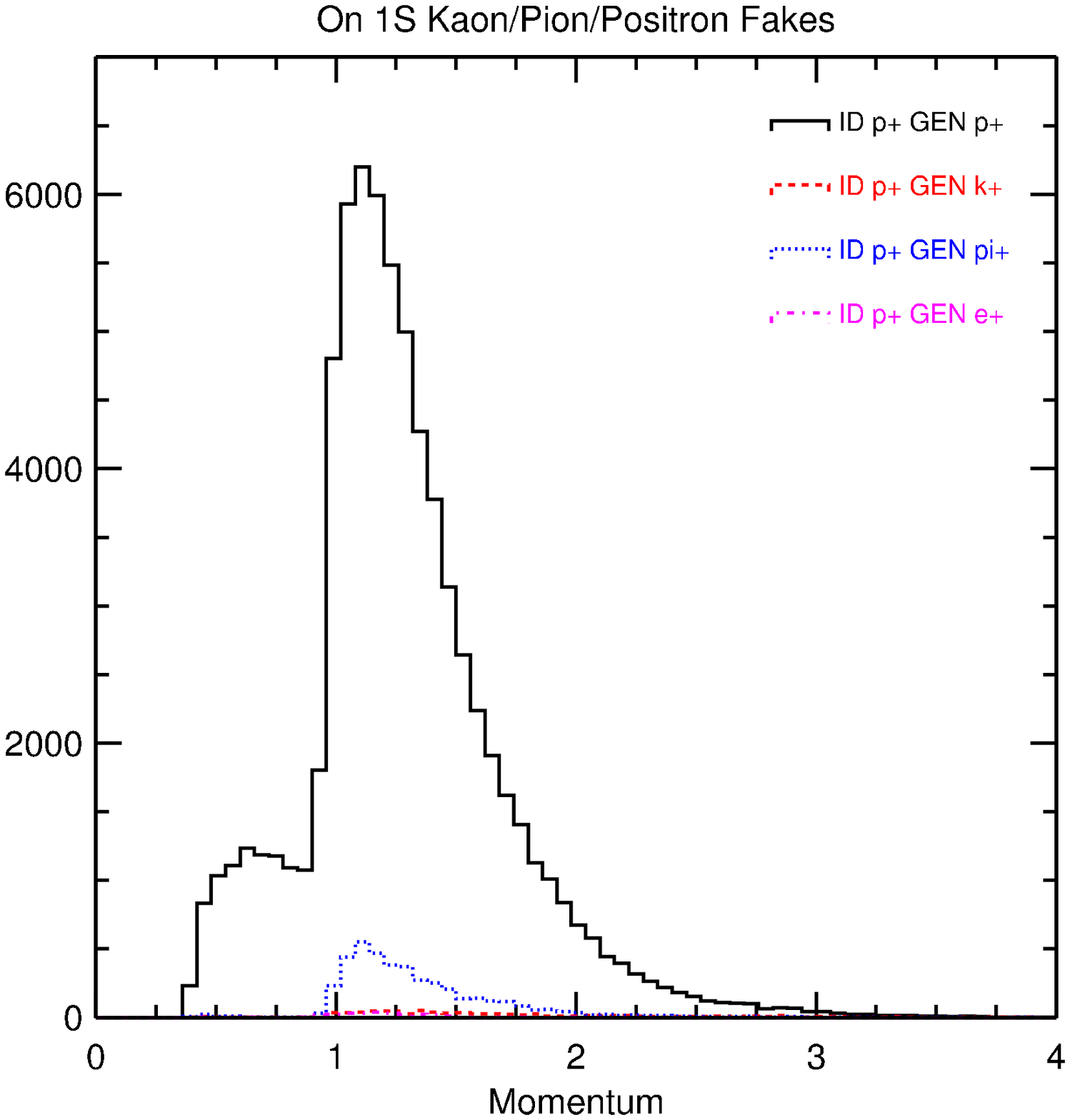}}
     \caption{\label{fig:b1sfakes} (Left) Proton fakes for a sample of below-1S
Monte Carlo simulations.  The solid black curve shows the number of all particles
identified as protons that were also tagged as true protons.  The red dashed 
(blue dotted, magenta dash-dot)
curve corresponds to those particles that were identified as protons, but that were
actually kaons (pions, positrons). (Right) Same for on-1S event simulations.}
\end{figure*}

We use Monte Carlo simulations to
assess fake proton backgrounds.
Figure~\ref{fig:b1sfakes} illustrates proton fakes for a sample of below-1S
Monte Carlo continuum simulations.  The solid black curve shows the number of all particles
identified as protons that were also tagged as true protons.  The red dashed 
(blue dotted, magenta dash-dot)
curve corresponds to those particles that were identified as protons, but that were
tagged at the generator level as true kaons (pions, positrons).  
Proton backgrounds are 
observed to be present at the $\sim$10\% level and are expected to largely
cancel in the ratio of ratios.

Note that, for all proton and antiproton identification, we require that
the minimum (anti-)proton momentum exceed 400 MeV to eliminate concerns
regarding protons ranging out in the beampipe.

\subsection{Signal Definition}
In this analysis we measure particle enhancements in 
both the ggg and gg$\gamma$
decays of the Upsilon system, relative to q\=q($\gamma$) production on the
underlying continuum.  
Our definition of enhancement is given quantitatively
as the continuum-subtracted resonance yield 
relative to the continuum yield.  
Thus defined, an enhancement of 1 indicates that
a given particle is produced as often (per event) 
on the continuum as on the resonance.
Note that our definition of `continuum' here means both
below-resonance continuum, as well
as resonance$\to$q\=q through vacuum-polarization;
i.e., all $e^+e^-\to$q\=q-like processes which must be
explicitly subtracted in determining the
characteristics of 3-gluon resonant decays.\footnote{Vacuum polarization
processes are subtracted by direct scaling of the continuum using
the $\Upsilon\to\gamma^\star\to q{\overline q}$ values tabulated
previously\cite{r:shawn}.}
Furthermore, note that
for the $\Upsilon$(2S) and $\Upsilon$(3S) data, there is no subtraction
of cascades to lower $\Upsilon$ states or $\chi_b$ decays. Therefore,
in what follows, 
``$\Upsilon$(2S)'' denotes a sum over $\Upsilon$(2S) direct, 
$\Upsilon(2S)\to\Upsilon$(1S)+X and $\Upsilon$(2S)$\to\chi_b$.
Assuming the direct decays of the $\Upsilon$ resonances are identical,
an $\Upsilon$(2S) enhancement smaller than that of the 
$\Upsilon$(1S) implies that the enhancements 
from the first and third processes
enumerated above are therefore smaller than for the $\Upsilon$(1S).

In general we have two
continuum-subtraction options:  we may determine enhancements for all resonances 
relative to the below-4S continuum (for which the statistics are largest, but the extrapolation
in energy and run period is also largest)
or we may find enhancements relative
to their individual 
below-resonant continua using data not more than
50 MeV away from the resonance of interest.  For mass-fitted particles
we normalize exclusively to the below-4S contiuum, as the individual continuua
(below-1S, -2S, and -3S) do not have high enough 
statistics to yield well-fitted
mass peaks.  For particle counts determined by the momentum spectra (protons
and antiprotons), 
we normalize to both the below-4S continuum as well as the
resonance-specific continuua 
and present the differences in the enhancements 
produced by the two as a systematic
error.

\subsection{Particle production in three-gluon vs.
q\=q events}
The previous CLEO-I analysis already observed significant enhancements of
protons and lambda's produced in 3-gluon decays
of the 1S relative to the below-4S continuum\cite{r:cleo84}. 
We repeat that analysis with our enlarged data set, as described
below.

Errors on particle yields
are obtained from the error on the fit if the particle count is obtained by 
fitting a mass peak ($\Lambda, \phi, f_2$) or by the square root of the total count
if the particle count is obtained from the momentum spectrum (p,$\overline{p}$).
For the ggg analysis, we determine enhancements as a function of
scaled momentum and also calculate
momentum-integrated enhancements for each particle, to allow comparison
with previous results.

\subsection{gg$\gamma$ Analysis}

For the gg$\gamma$ analysis we normalize 
the particle counts in a given bin to the
photon counts in the same bin.  
In this case the bins are determined by the photon
momentum and not the particle momentum (as is the case in the ggg analysis).  For each bin,
we then find the fractional contamination of resonance photons due to the underlying
continuum.  Once this fractional contamination is known, the resonance
yield can be extracted by straight-forward algebra. As a check of the 
numerical procedure, we have verified that the enhancements obtained
for $\Upsilon$(4S) (which should produce no photons with $x_\gamma>$0.5,
since its width is dominated by B\=B) relative to the below-4S continuum
are consistent with zero.

\begin{equation}
Fraction_{Continuum}=\frac{\sigma_{x_\gamma>0.5}^{Continuum}}{\sigma_{x_\gamma>0.5}^{Resonance}}\left(\frac{E_{beam}^{Continuum}}{E_{beam}^{Resonance}}\right)^2
\label{eq:cntfrac}
\end{equation}

\section{Results}

\subsection{ggg Enhancements}

\subsubsection{Baryon Enhancements}

\begin{figure*}
     \includegraphics[width=3.5in,height=3.5in]{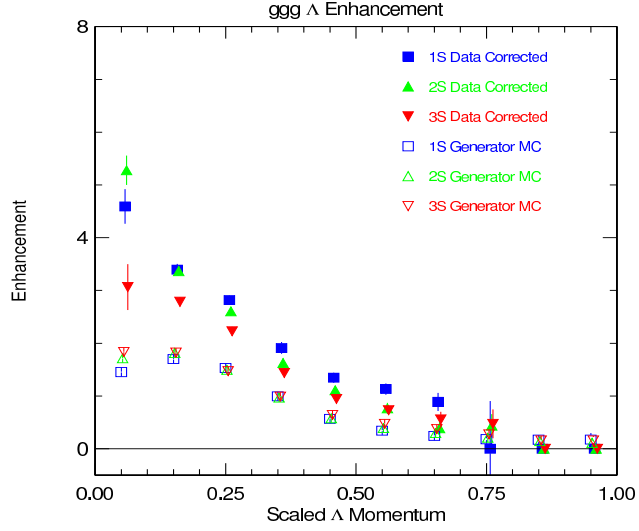}
     \caption{\label{fig:ggglam} Raw
enhancements for ggg$\to\Lambda+X$ binned
according to scaled momentum ($p_\Lambda/E_{beam}$). 
Blue square (green triangle, red diamond) symbols correspond to 
enhancements on the 1S (2S, 3S) resonance.  Closed symbols are data, 
open symbols are JETSET Monte Carlo. }
\end{figure*}

Figure~\ref{fig:ggglam} presents our $\Lambda$ enhancements binned according to scaled
momentum, that is the momentum of the particle divided by the beam energy.
In the figure, blue square (green triangle, red diamond) symbols correspond to enhancements on the 1S
(2S, 3S) resonance.  Closed symbols 
are data and open symbols are JETSET 7.4\cite{JETSET7.4}
event generator simulations followed by the full CLEOIII
GEANT-based\cite{r:GEANT} Monte Carlo detector simulation. We note that the
acceptance in the lowest momentum bin is very sensitive to 
absorption of protons in beam-pipe.

\begin{figure*}
\flushleft{\includegraphics[width=3.5in,height=3.5in]{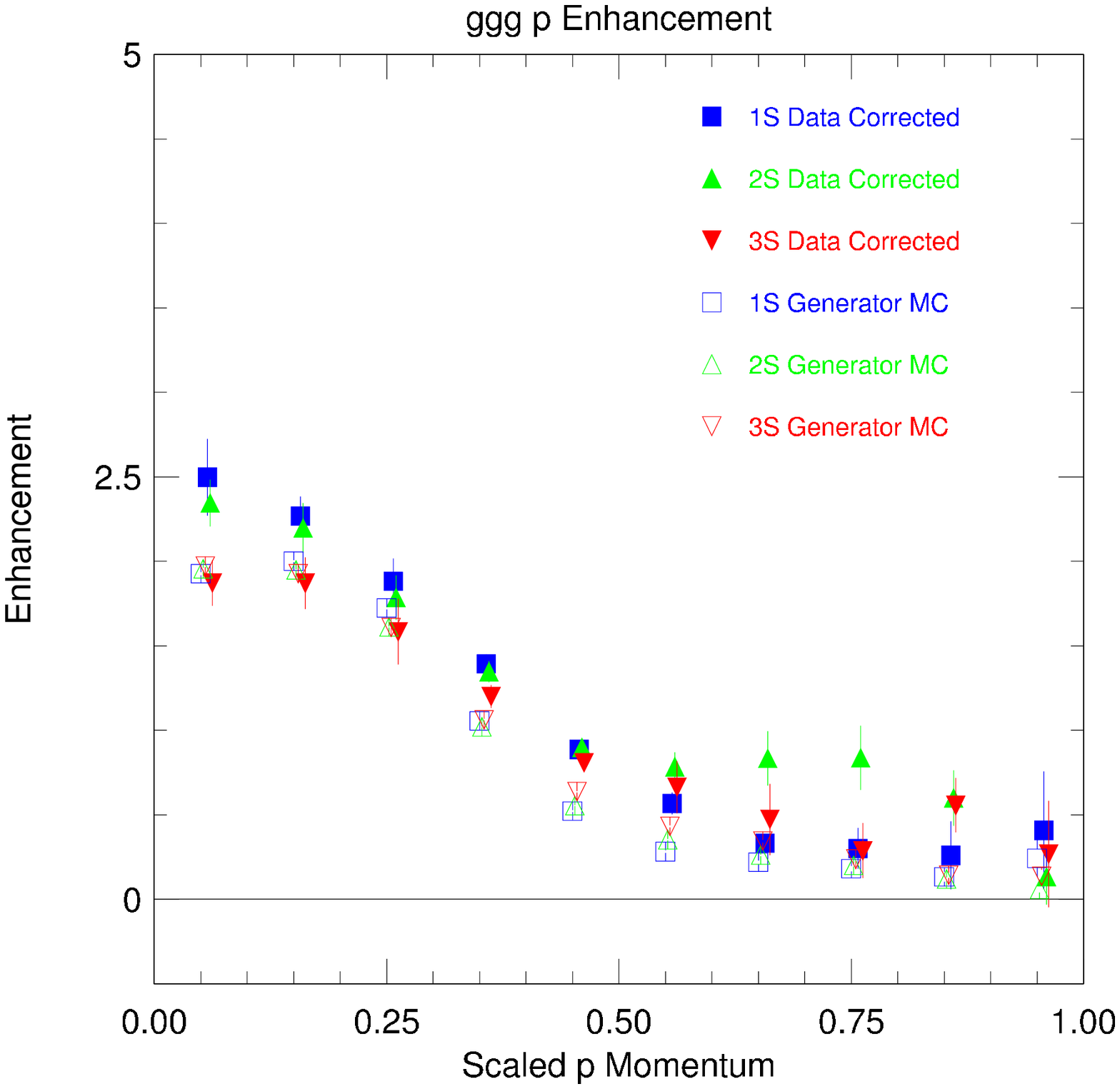}}
\vspace{-3.65in}
\flushright{\includegraphics[width=3.5in,height=3.5in]{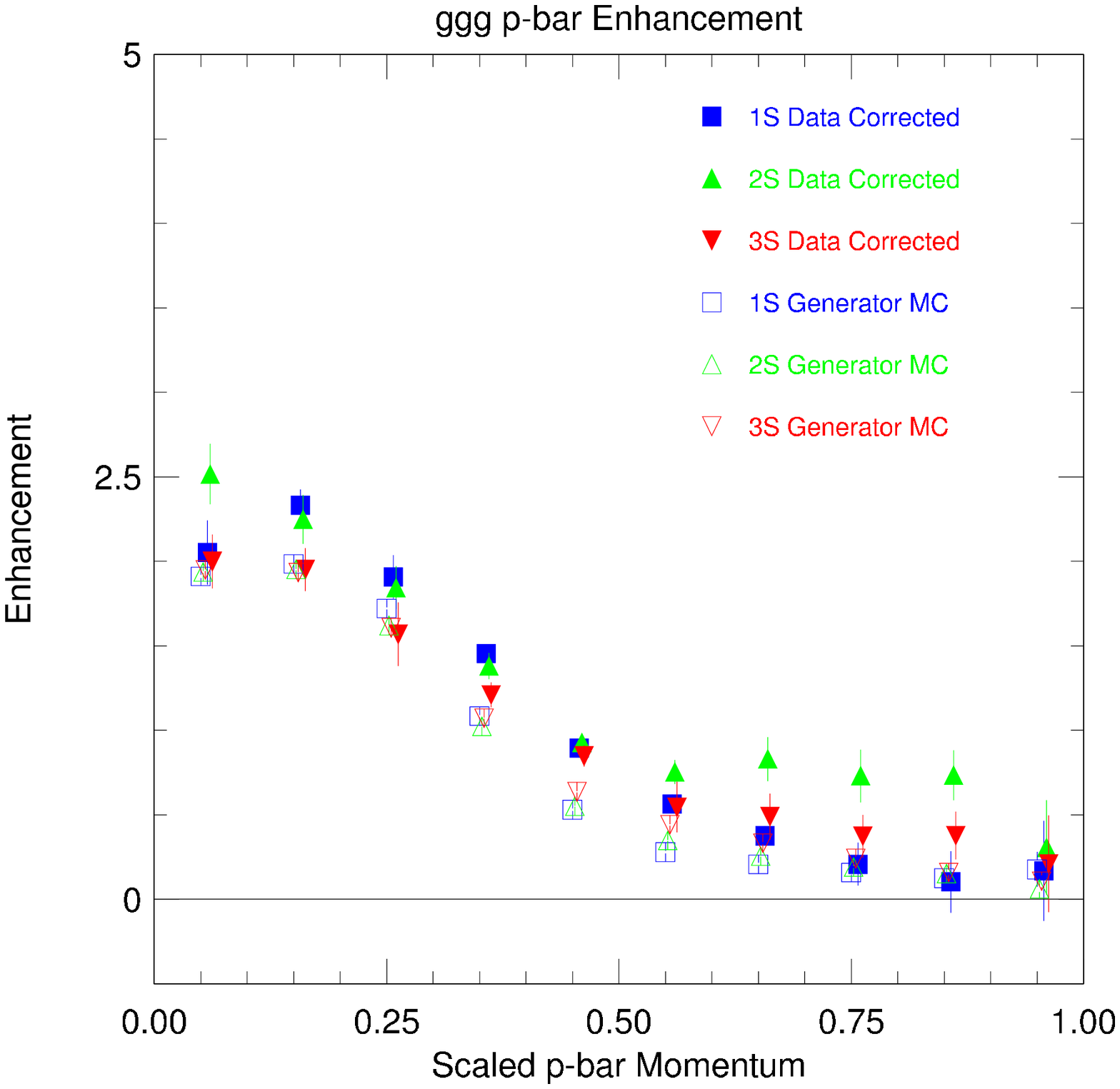}}
     \caption{\label{fig:gggpro} (Left) 
Raw enhancements for $ggg\to p+X$ binned
according to scaled momentum.  Statistics of continuum subtraction
are largest in lower momentum bins.
Blue square (green triangle, red diamond) symbols correspond to 
enhancements on the 1S (2S, 3S) resonance.  Closed symbols are data, 
open symbols are JETSET Monte Carlo. (Right) Same for antiprotons.}
\end{figure*}

Figure~\ref{fig:gggpro} 
shows the proton and antiproton enhancements.
The consistency between the two indicates 
that beam-wall and beam-gas backgrounds are not substantial.

\subsubsection{$\phi$ and $f_2$(1270) Enhancements}

\begin{figure*}
\flushleft{\includegraphics[width=3.5in,height=3.5in]{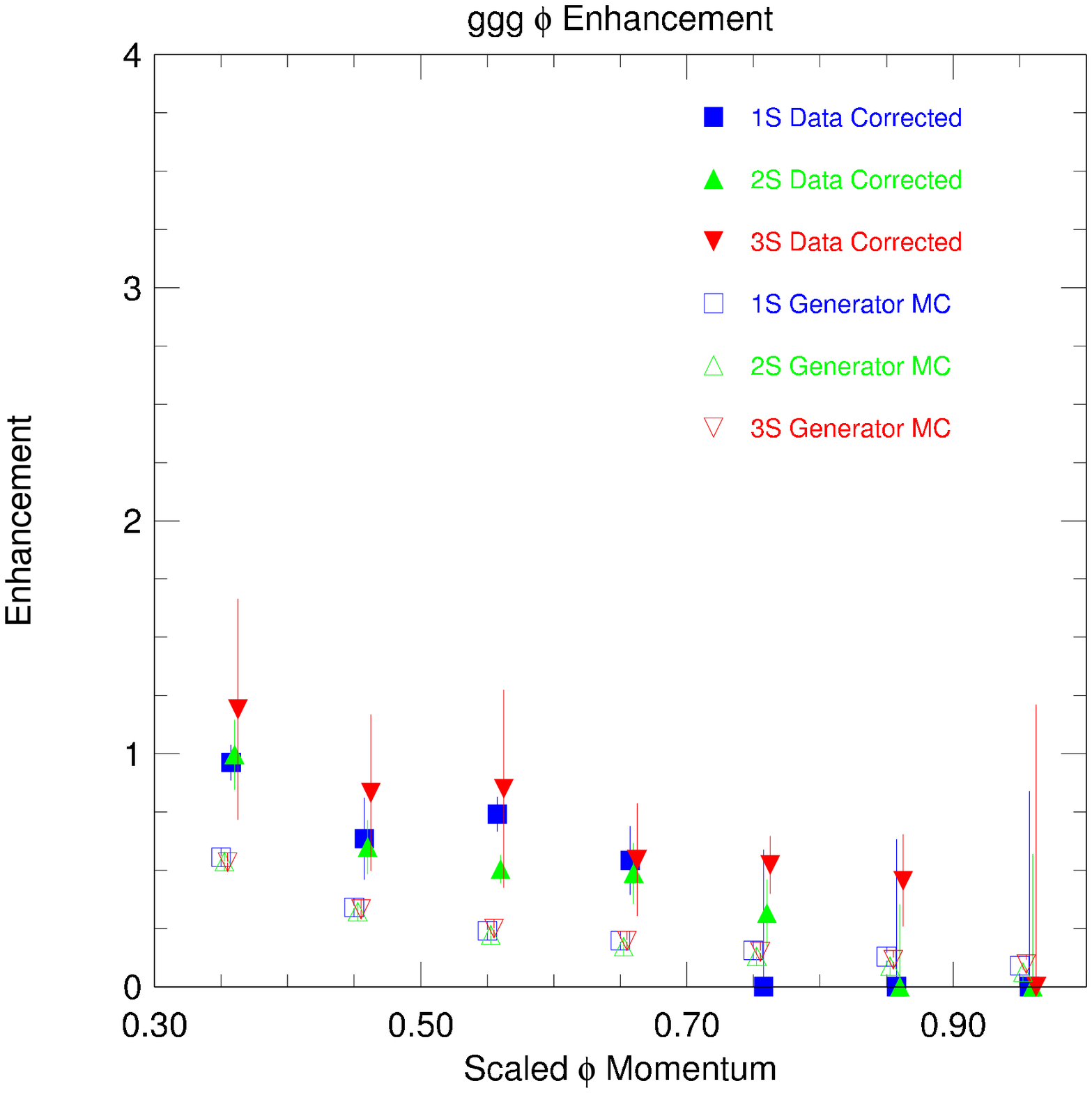}}
\vspace{-3.65in}
\flushright{\includegraphics[width=3.5in,height=3.5in]{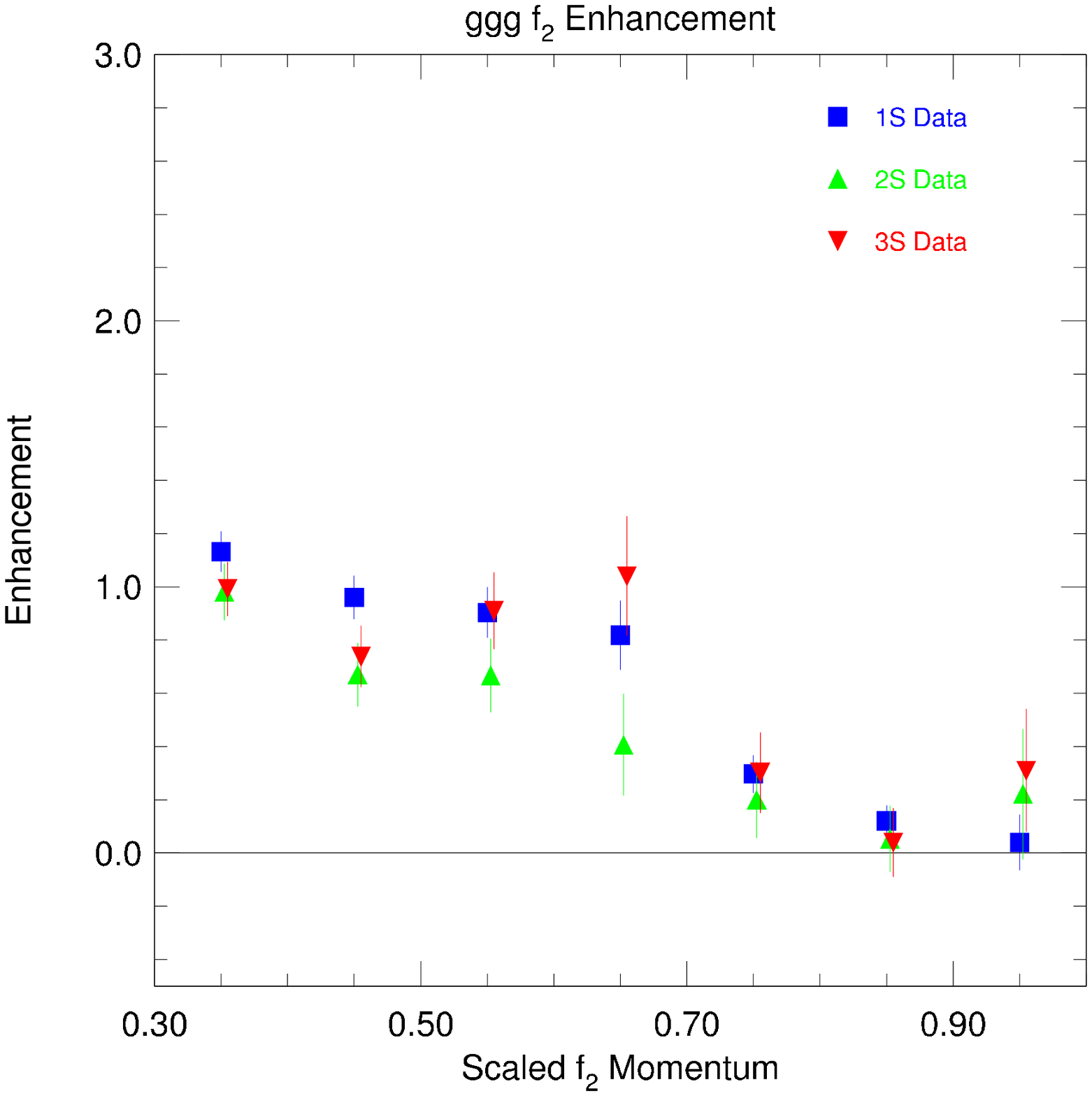}}
     \caption{\label{fig:gggphi} (Left) Raw
enhancements for ggg $\phi$ binned
according to scaled momentum.  Blue square (green triangle, red diamond) symbols correspond to 
enhancements on the 1S (2S, 3S) resonance.  Closed symbols are data, 
open symbols are JETSET Monte Carlo. 
(Right) Enhancements for $f_2$(1270).}
\end{figure*}

Figure~\ref{fig:gggphi} shows $\phi$ enhancement results binned
according to scaled momentum.  Symobls are as above with blue square (green triangle, red diamond)
 symbols correspond to enhancements on the 1S (2S, 3S) 
resonance.  Closed symbols are data and open symbols are JETSET Monte Carlo.
Here, we have normalized
the resonant production at 9.46 GeV to the continuum production at
10.55 GeV. Therefore, for $\phi$ production,
the lowest momentum bins for the resonance
are particularly sensitive to low-momentum kaon acceptance. 
Figure~\ref{fig:gggphi} also shows the f$_2$ enhancement results binned
according to scaled momentum.  The f$_2$ peak is not well-defined 
at low momentum (lowest two bins).  No Monte Carlo comparison
is presented since
our current Monte Carlo, by default, will not generate tensor particles.

\subsubsection{Particle Momentum-Integrated Enhancements}
\begin{figure*}
     \includegraphics[width=5in,height=5in]{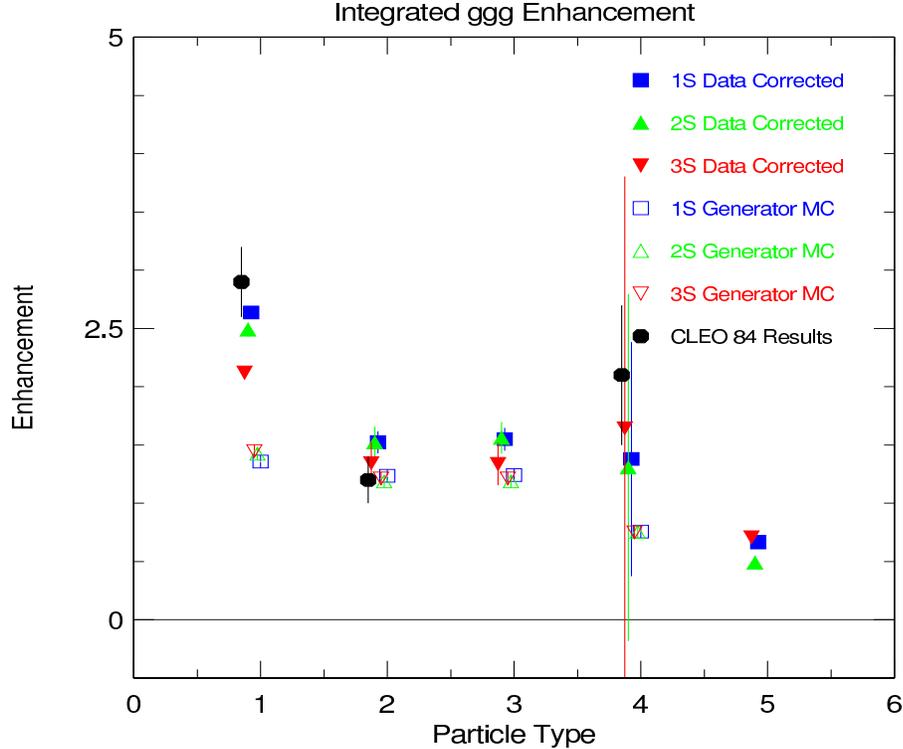}
     \caption{\label{fig:ggginteg} Compilation of momentum-integrated 
enhancements for $ggg$ events.  Blue square (green triangle, red diamond) symbols correspond to 
enhancements on the 1S (2S, 3S) resonance.  Closed symbols are data, 
open symbols are JETSET Monte Carlo. The x-axis is arbitrary, chosen
to display all particles on one plot: $\Lambda$ is plotted at x=1,
protons at x=2, $\overline{p}$ at x=3, $\phi$ at x=4, and f$_2$ at x=5. 
Systematic errors and relative efficiencies have now been included
for this compilation. The CLEO84 study did not measure an
enhancement for $f_2(1270)$ and also only presented
a single enhancement for the sum of protons and antiprotons.}
\end{figure*}

Figure~\ref{fig:ggginteg} shows the particle momentum-integrated enhancements
for each particle.  We note that the baryons
($\Lambda$, p, $\overline{p}$) have enhancements greater than 1,
while the mesons ($\phi$, f$_2$) have enhancements less than 1.
Our results are, in general, numerically consistent with the
prior CLEO-I analysis, albeit with considerably higher
statistics.

\subsection{gg$\gamma$ Enhancements}

There are
sufficient statistics
to present binned enhancements (in this case binning is done according
to tagged photon momentum, not according to individual particle momentum) 
for $\Lambda$, protons
and $\overline{p}$. For all particles, we also present
momentum-integrated enhancements.

\subsubsection{Baryon Enhancements}

\begin{figure*}
     \includegraphics[width=3.5in,height=3.5in]{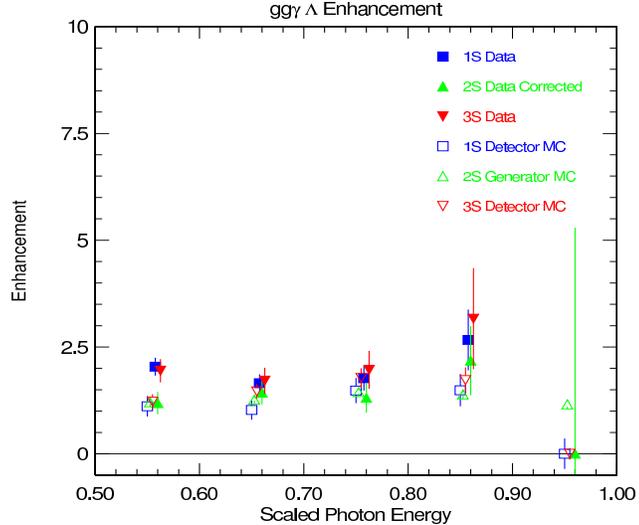}
     \caption{\label{fig:gggamlam} 
Raw enhancements for gg$\gamma\to\Lambda+X$ 
binned
according to scaled photon momentum.  
We have integrated over all $\Lambda$ momenta to make these plots, in 
events having a photon with the indicated x-value.
Blue square (green triangle, red diamond) symbols correspond to 
enhancements on the 1S (2S, 3S) resonance.  Closed symbols are data, 
open symbols are JETSET Monte Carlo. }
\end{figure*}

Figure~\ref{fig:gggamlam} shows $\Lambda$ results binned according to scaled
photon momentum, with the color scheme as before. 
Errors are large in the highest bin due to limited statistics.

\begin{figure*}
\flushleft{\includegraphics[width=3.5in,height=3.5in]{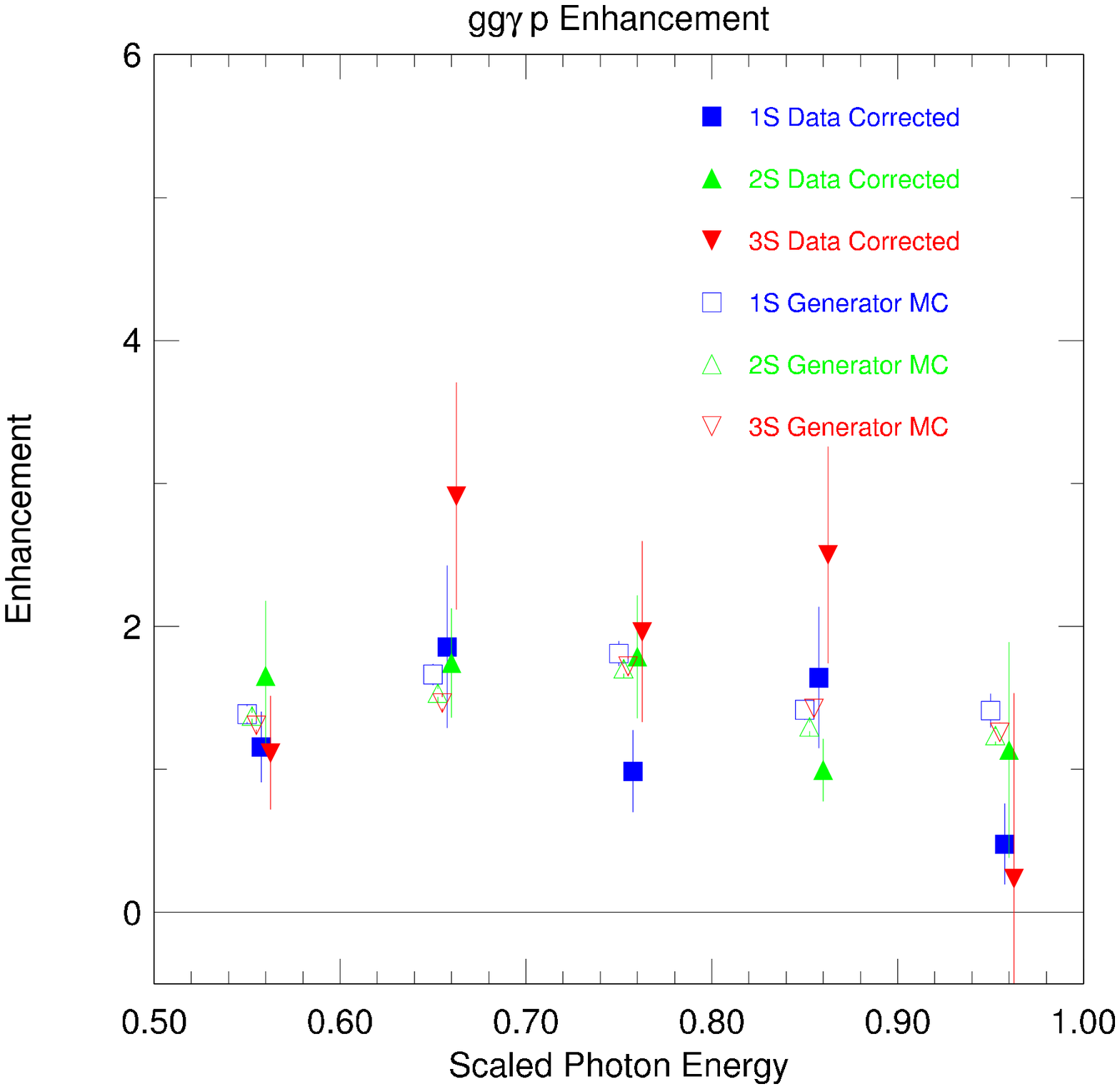}}
\vspace{-3.65in}
\flushright{\includegraphics[width=3.5in,height=3.5in]{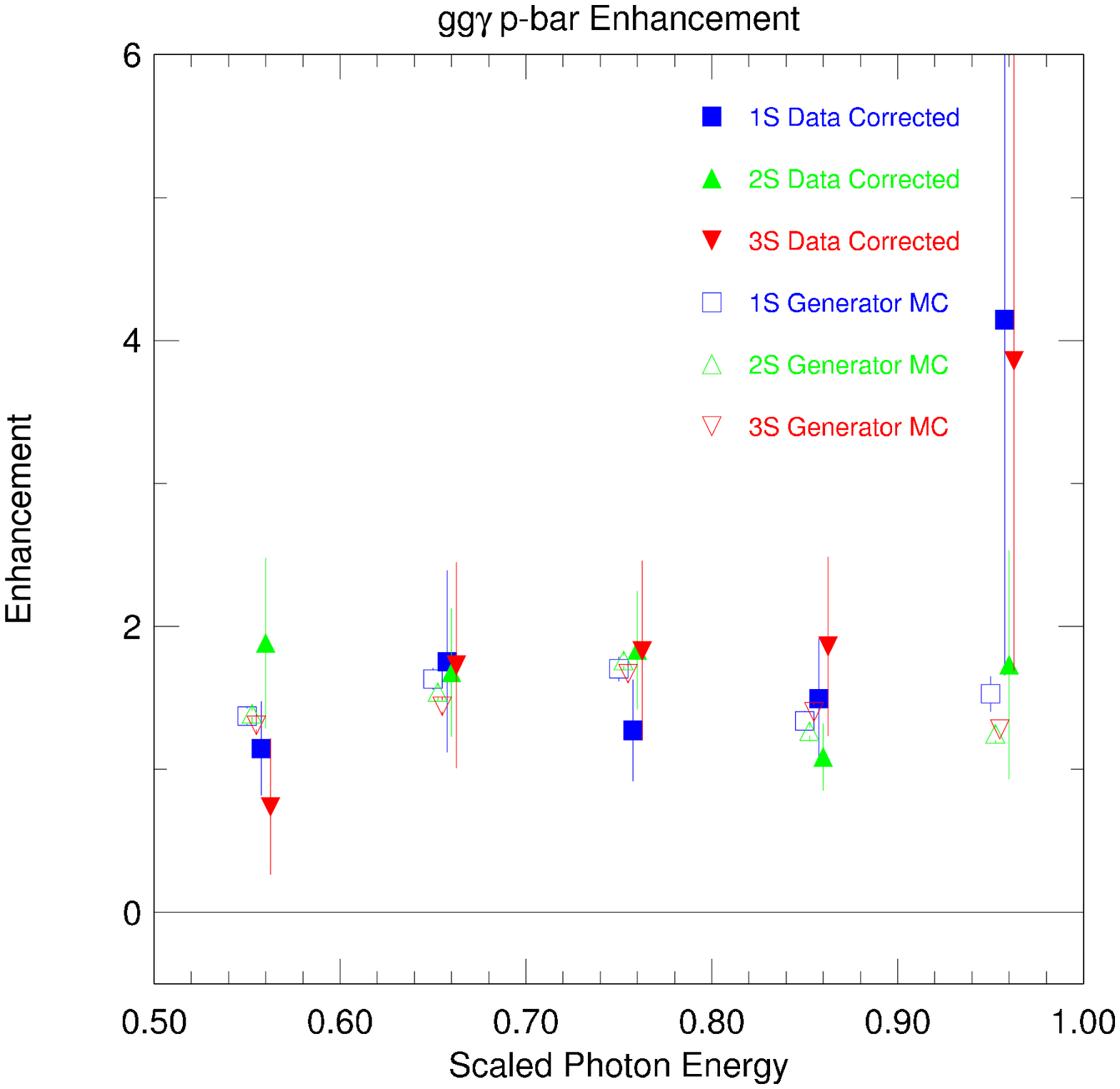}}
     \caption{\label{fig:gggampro} (Left) Raw
enhancements for gg$\gamma$ p binned
according to photon momentum.  
We have integrated over all proton momenta to make these plots, in 
events having a photon with the indicated x-value.
Blue square (green triangle, red diamond) symbols correspond to 
enhancements on the 1S (2S, 3S) resonance.  Closed symbols are data, 
open symbols are JETSET Monte Carlo. (Right) Same for antiprotons. }
\end{figure*}

Figure~\ref{fig:gggampro} shows proton and antiproton 
enhancement results binned according
to scaled photon momentum. We immediately note somewhat smaller
enhancements compared to the 3-gluon enhancements presented previously.

\subsubsection{Momentum-Integrated Enhancements}
\begin{figure*}
     \includegraphics[width=5in,height=5in]{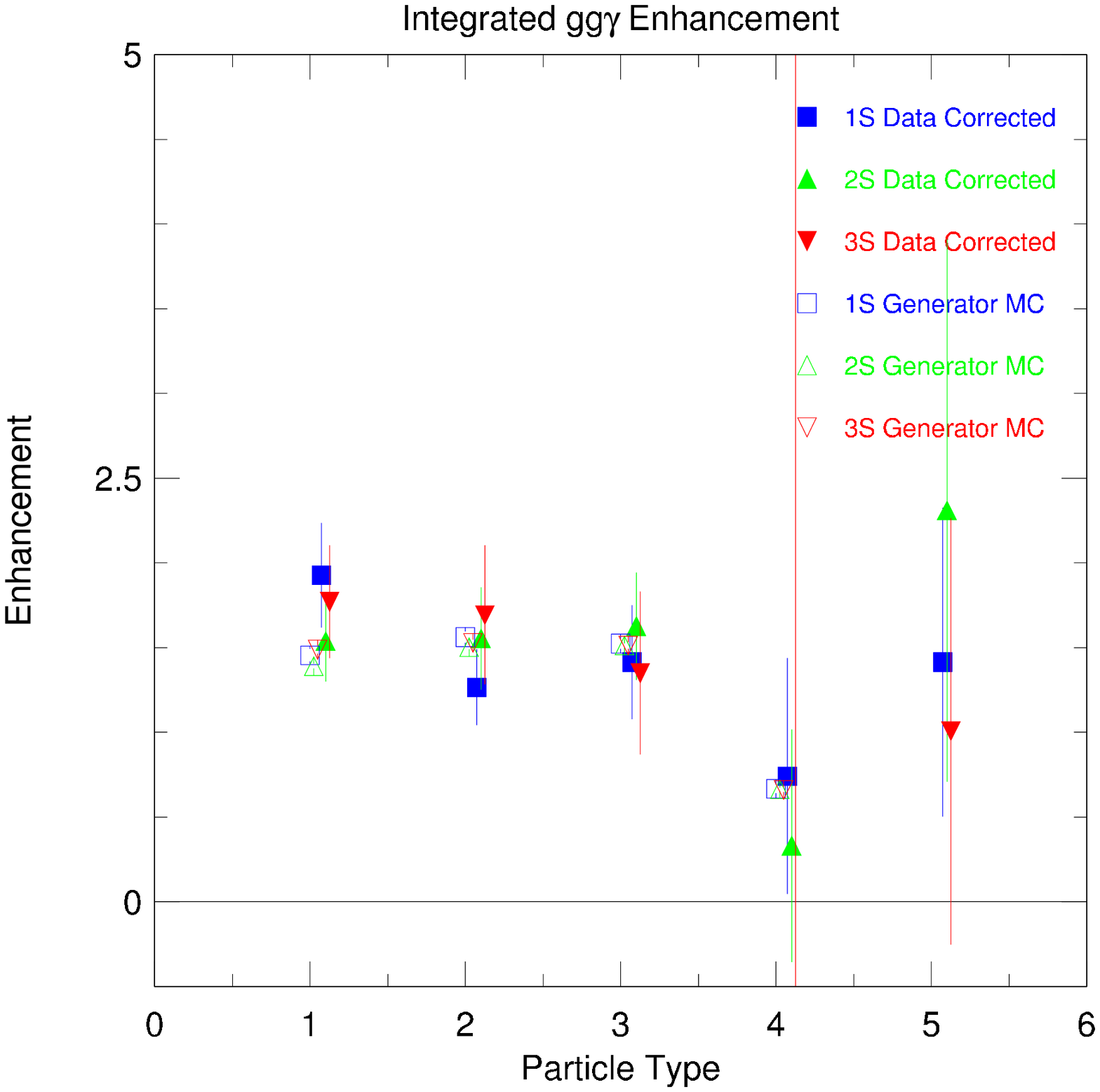}
     \caption{\label{fig:gggaminteg} Momentum-integrated 
enhancements for gg$\gamma$.  Blue square (green triangle, red diamond) symbols correspond to 
enhancements on the 1S (2S, 3S) resonance.  Closed symbols are data, 
open symbols are JETSET Monte Carlo. The x-axis is arbitrary, chosen
to display all particles on one plot: $\Lambda$ is plotted at x=1,
p at x=2, $\overline{p}$ at x=3, $\phi$ at x=4, and f$_2$ at x=5. 
Systematic errors have been included and relative efficiency corrections
have been applied.}
\end{figure*}

Figure~\ref{fig:gggaminteg} shows the momentum-integrated enhancements
for each particle.  We note that the enhancements are considerably
mitigated in comparison with the case of three-gluon fragmentation.

\section{$\chi_{b,J}\to$p+X}
Photon transitions of the $\Upsilon$(2S) and 
$\Upsilon$(3S) to the $\chi_b$ states allow
us to measure the baryon yields, in association with a radiative
transition photon `tag'. Typical photon tag energies in this
case are of order 80-160 MeV. Due to the large background in this
region and therefore limited
statistical power, the data permit only an extraction of the proton and antiproton
enhancements. Of particular interest is the proton yield
in $\chi_{b,2}$ vs. $\chi_{b,1}$ decays; the former is
expected to be dominated by decays via two-gluons, the latter
is expected to be dominated by decays to q\=q(g), with the gluon
expected to carry away very little momentum.

We first conduct a Monte Carlo study to determine the relative 
efficiency of reconstructing a J=2 transition
photon relative to J=1 event,
and also the efficiency when we require that a proton be found in
addition to the transition photon.
We compile statistics on the $\chi_b(')\to p({\overline p})+X$ 
analyses, separately for J=0/J=1 and for J=2/J=1. For the latter,
we fit the signal to a double Gaussian, so that the errors in 
the areas of the two
peaks are correlated and the statistical significance of the
signals enhanced. For the former, we simply fit two Gaussians directly.
We find that the efficiency for reconstructing a photon-proton
correlation in a two-gluon
decay is approximately 95\% that for a photon-proton correlation
in q\=q(g) event.

To check the sensitivity to our particle identification
criteria, we have compared results using
very tight proton identification 
requirements (with a reduction in efficiency by more than 50\%)
vs. the 'standard' loose proton identification
criteria used above. We obtain a comparable
correction factor using more restrictive particle identification criteria.

Results are presented in the Table \ref{tab:chib}. 
We note that the
observed enhancements are, again, smaller than those
observed in comparing three-gluon fragmentation from the $\Upsilon$ resonance
with q\=q fragmentation.

\begin{table}
\caption{Summary of inclusive proton (and antiproton) results for
$\chi_{b,J}$ decays. For checks of internal consistency,
data have been separated into sub-samples, labeled with capital
Roman letters. For J=2 relative to
J=1, e.g., the scale of systematic uncertainties is set by
the constancy of the value across datasets (r.m.s.$\sim$0.03),
the magnitude of relative efficiency corrections ($\sim$0.05)
and the consistency of results obtained using different particle
identification criteria.}
\label{tab:chib}
\begin{tabular}{|c|c|c|c|} \hline
Dataset & particle & ($J=2\to p+X$) & 
($J=0\to p+X$) \\
        & Identification & /($J=1\to p+X$) & /($J=1\to p+X$) \\ \hline
 
(3S A) & loose & $1.116\pm0.017$ & $1.19\pm0.046$ \\ 
 
(3S B) & loose & $1.080\pm0.016$ & $1.00\pm0.034$ \\ 
 
(3S C) & loose & 

$1.086\pm0.011$ &

$1.054\pm0.047$ \\ 
 
(3S D) & tight & 

$1.103\pm0.027$ &

$1.091\pm0.097$ \\ \hline
3S, all & & $1.11\pm0.01\pm0.04$ & $1.092\pm0.03\pm0.06$ \\ \hline

(2S A) & tight & 1.066$\pm$0.028 & 1.03$\pm$0.13 \\

(2S B) & loose & 1.075$\pm$0.018 & 

$1.36\pm0.15$ \\

(2S C) & loose & 1.076$\pm$0.017 & 

0.99$\pm$0.11 \\

(2S D) & loose & 1.065$\pm$0.015 & 

1.06$\pm$0.11 \\ \hline

(2S B) & tight & 1.076$\pm$0.047 & 

$1.39\pm0.28$ \\

(2S C) & tight & 1.039$\pm$0.040 & 

1.17$\pm$0.22 \\

(2S D) & tight & 1.024$\pm$0.035 & 

0.88$\pm$0.20 \\ \hline
2S, all & & $1.068\pm0.010\pm0.04$ & $1.11\pm0.15\pm0.20$ \\ \hline

Monte Carlo (3S A) & loose & $1.057\pm0.016$ & $1.030\pm0.072$ \\

Monte Carlo (3S A) & tight & $1.034\pm0.015$ & $1.042\pm0.066$ \\

Monte Carlo (3S B) & tight & $1.041\pm0.013$ & $1.051\pm0.049$ \\ \hline
MC, 3S all & & $1.043\pm0.01$ & $1.043\pm0.036$ \\ \hline

Monte Carlo (2S A) & tight & 1.052$\pm$0.014 & 1.121$\pm$0.058 \\

Monte Carlo (2S A) & loose & 1.043$\pm$0.015 & 1.076$\pm$0.061 \\ \hline
MC, 2S all & & $1.046\pm0.01$ & $1.061\pm0.025$ \\ \hline
\end{tabular}
\end{table}

\message{In Monte Carlo, there is, 
therefore no ``gluonic baryon enhancement'' as measured relative to
non-protons --
the ratio of 
$((\chi_{b,2}\to p+X/\chi_{b,1}\to p+X)/
(\chi_{b,2}\to {\not p}+X/\chi_{b,1}\to {\not p}+X))$=1.004
is statistically consistent with unity, as is the corresponding
ratio for 
$((\chi_{b,0}\to p+X/\chi_{b,1}\to p+X)/
(\chi_{b,0}\to {\not p}+X/\chi_{b,1}\to {\not p}+X))$ (=0.987).
In data, the corresponding ratio is:
$((\chi_{b,2}\to p+X/\chi_{b,1}\to p+X)/
(\chi_{b,2}\to {\not p}+X/\chi_{b,1}\to {\not p}+X))$=0.993/0.981.}

\section{Cross-Checks and Systematics}
\begin{figure*}
\flushleft{\includegraphics[width=3.5in,height=3.5in]{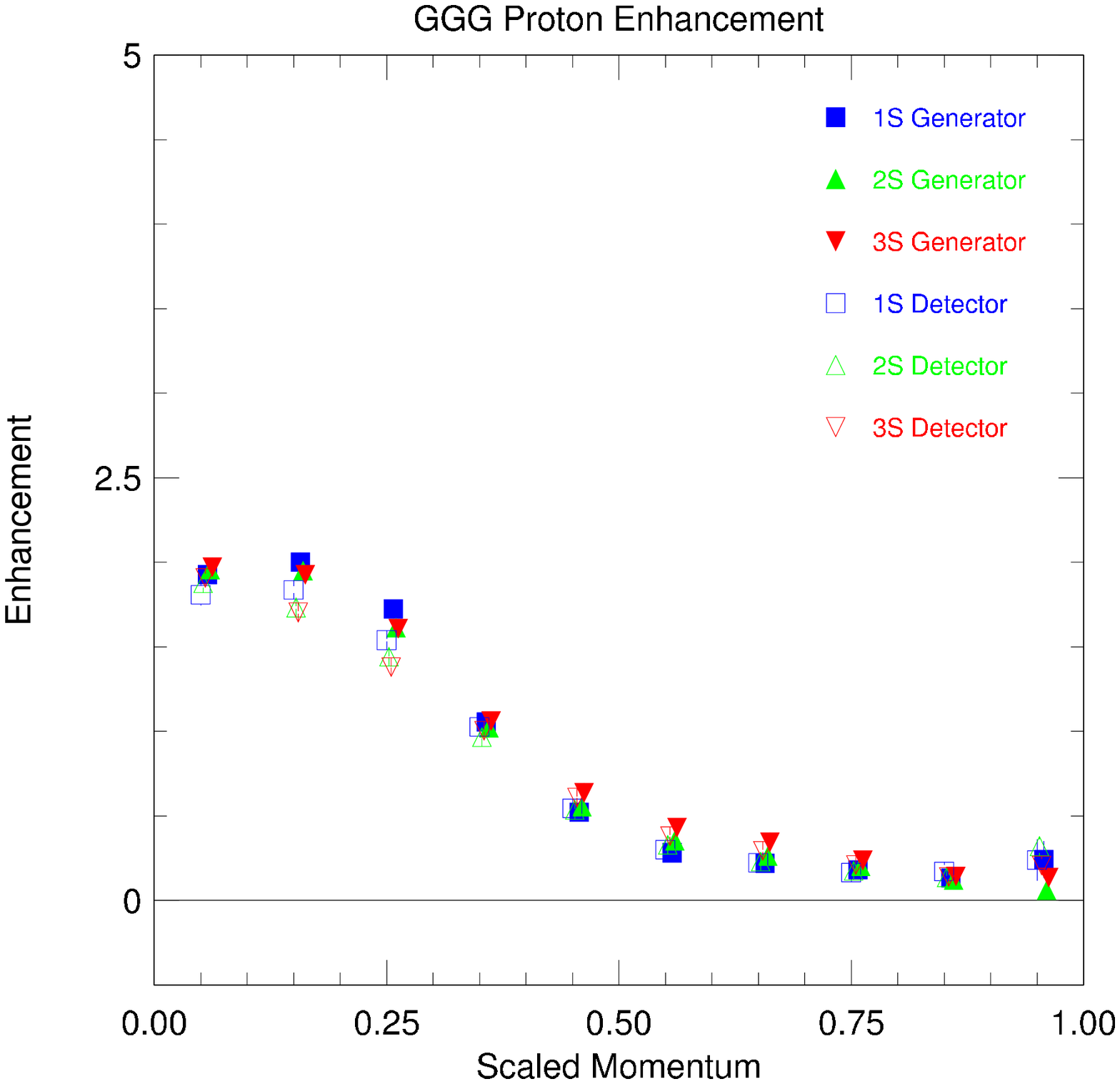}}
\vspace{-3.65in}
\flushright{\includegraphics[width=3.5in,height=3.5in]{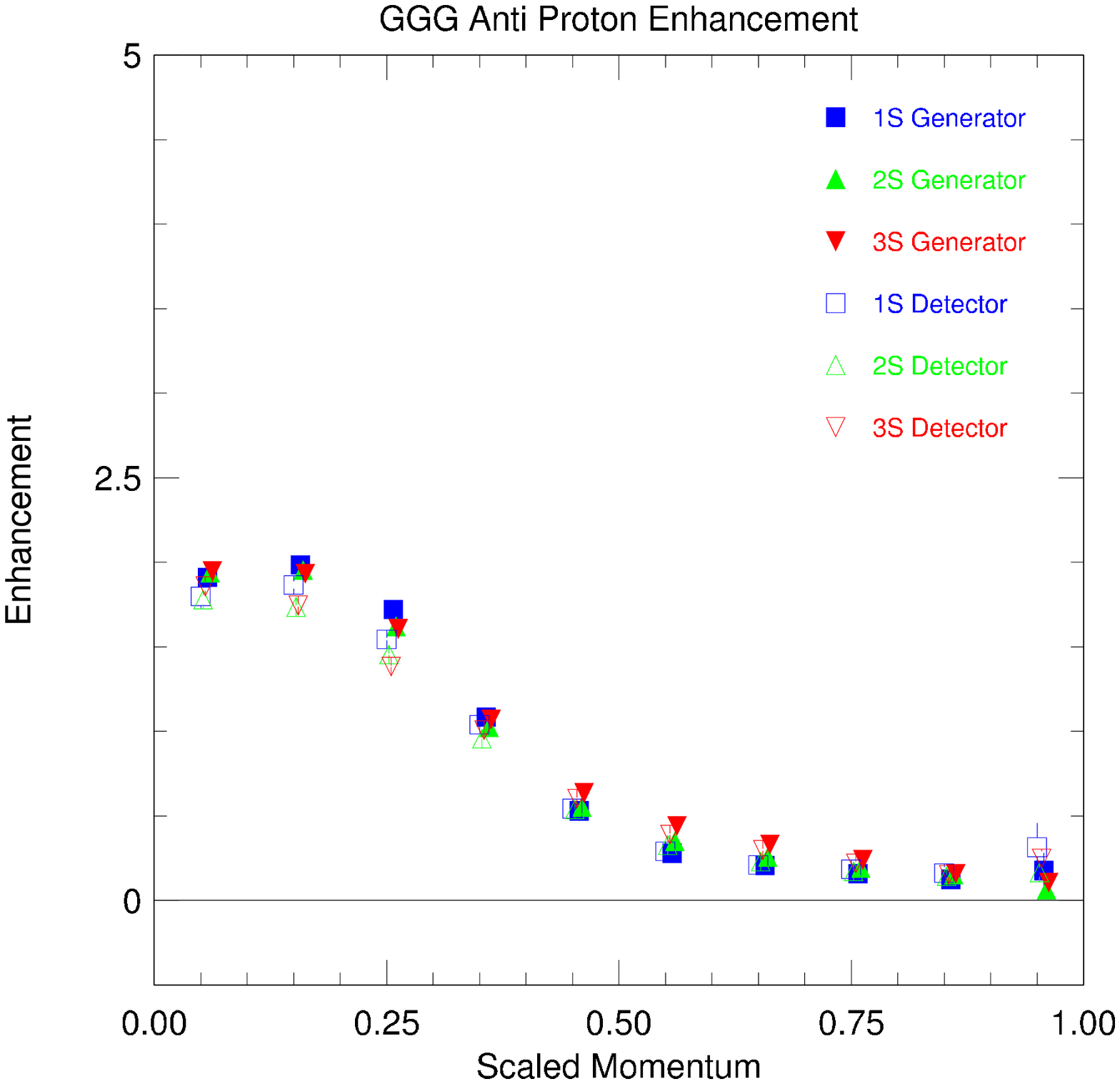}}
     \caption{\label{fig:gggprodg} (Left) Scaled momentum binned 
enhancements for $ggg\to p+X$.  Blue square (green triangle, red diamond) symbols 
correspond to enhancements on the 1S (2S, 3S) resonance.  Closed (open)
symbols are generator (detector) level Monte Carlo enhancements. (Right)
Same for antiprotons. }
\end{figure*}

\begin{figure*}
\flushleft{\includegraphics[width=3.5in,height=3.5in]{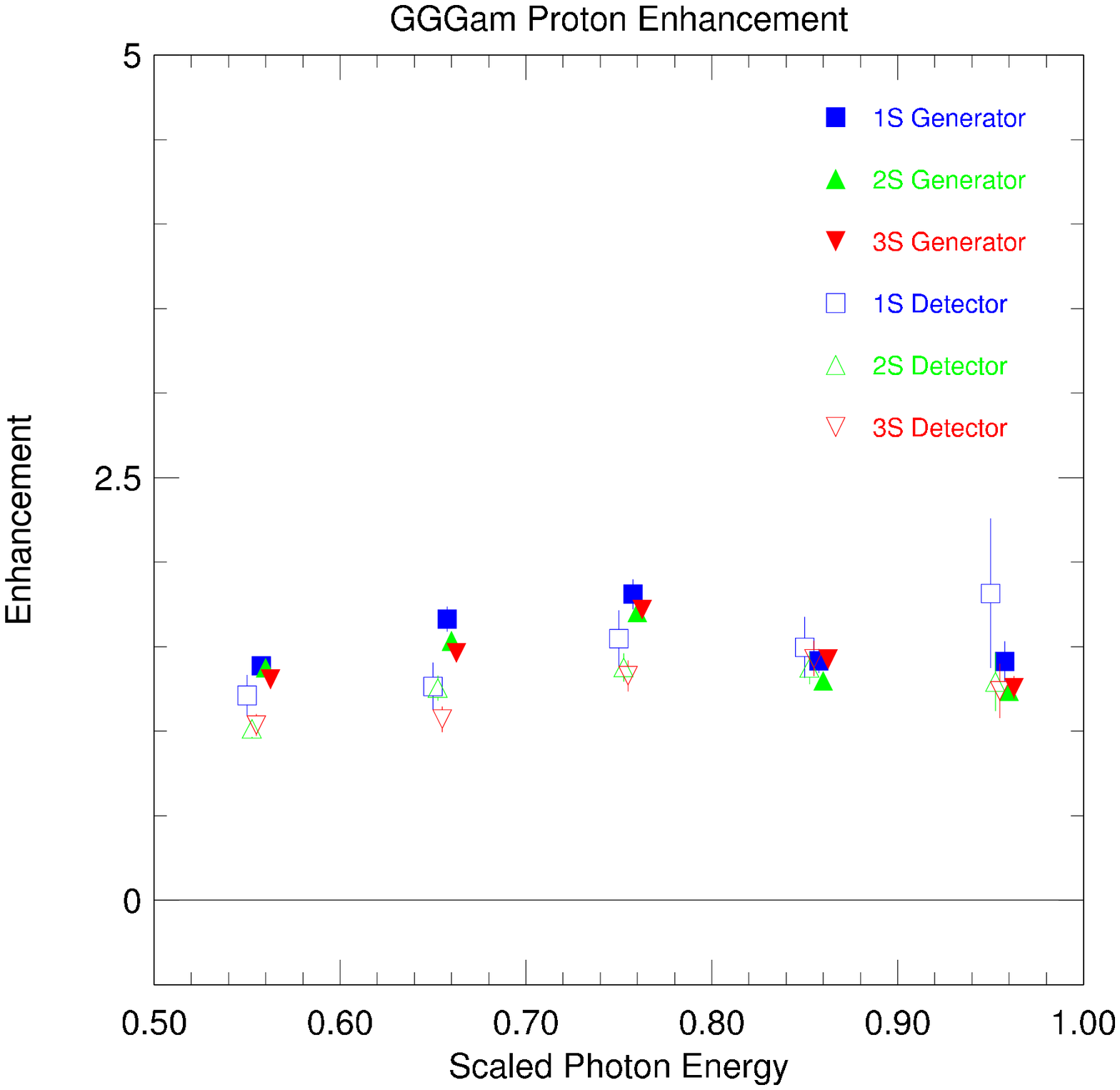}}
\vspace{-3.65in}
\flushright{\includegraphics[width=3.5in,height=3.5in]{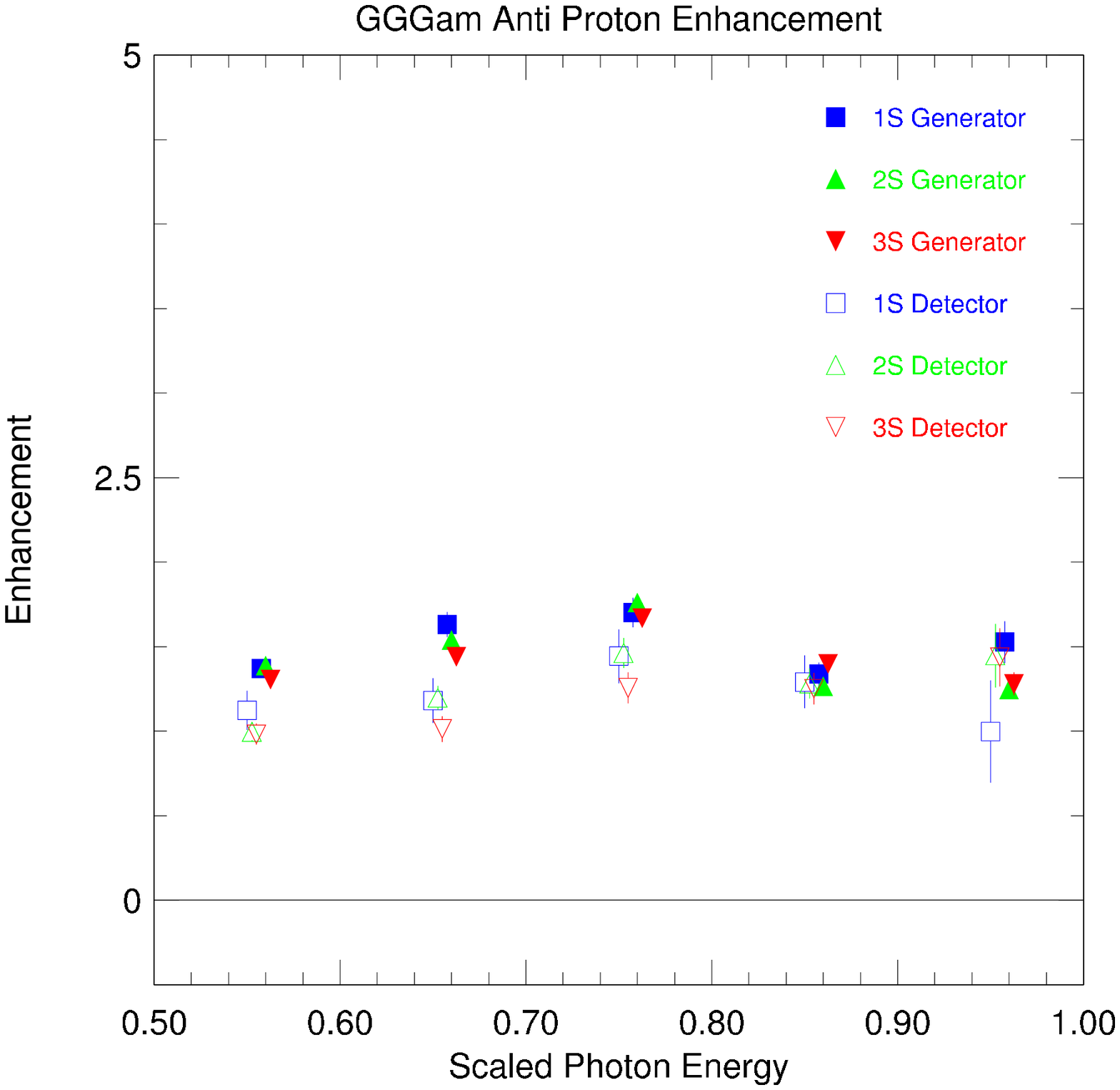}}
     \caption{\label{fig:gggamprodg} (Left) Scaled momentum binned 
enhancements for gg$\gamma$ decays to p.  Blue square (green triangle, red diamond) symbols 
correspond to enhancements on the 1S (2S, 3S) resonance.  Closed (open)
symbols are generator (detector) level Monte Carlo enhancements. (Right)
Same for antiprotons. }
\end{figure*}

In order to verify our procedures and probe possible
systematic uncertainties, two primary cross-checks were employed.
We first compare the Monte Carlo enhancements at the
event generator-level with those determined after the generated
events are processed through the full
CLEO-III detector simulation (``detector-level''), as a function of momentum.
In general, these enhancements will vary for several reasons, including
differences in:
a) the efficiencies for finding recoil particles in q\=q$\gamma$ 
vs. gg$\gamma$ events resulting from angular distribution,
event multiplicity, and particle momentum differences,
b) event selection efficiencies, c) $\pi^0$
contamination levels, and d) recoil center-of-mass
discrepancies, e.g., between the continuum data under
the $\Upsilon$(1S) resonance vs. the below-$\Upsilon$(4S)
continuum.
In cases where the generator-level and detector-level
enhancements are statistically inconsistent with
each at the $2\sigma$ level,
we use the ratio between the generator-level
and detector-level enhancements as a correction factor and
conservatively take half of the amount by which this
correction deviates from unity as an estimated systematic error.
(Note that these corrections have already been incorporated
into the results presented in Figures and \ref{fig:ggginteg}
and \ref{fig:gggaminteg}). Figures \ref{fig:gggprodg} 
and \ref{fig:gggamprodg} shows the comparison of proton
enhancements determined at the event-generator
vs. post-detector-simulation levels of Monte Carlo simulation.
Integrated over momentum, typical
corrections are typically of order 10\%.

To test the sensitivity of our analysis procedures across
different running periods, we have calculated the enhancements
for photon-tagged $\Upsilon$(4S) on-resonance events vs.
photon-tagged below-$\Upsilon$(4S) continuum events, spanning
the full CLEO-III data set. Since $\Upsilon$(4S)$\to$B\=B$\sim$100\%,
we expect that any event having a photon with $x_\gamma>0.5$ is
a continuum event. Hence, the calculated enhancement should be
zero. In all cases, save for antiprotons, we find
good agreeement between the below-4S continuum particle 
yields per photon tag, and the on-4S particle
yields per photon tag. For antiprotons,
we find deviations from the null expectation at the
level of $\approx$10\%, and incorporate these deviations
(bin-by-bin in momentum) into our total systematic error
for that particular case.

\section{Discussion and Summary}
We compare, for the first time, particle production in
two-gluon vs. quark-antiquark fragmentation. We find that,
in particular, baryon production (per event) in two-gluon decays is 
somewhat
smaller than that observed in three-gluon decays, particularly
in the case of $\Lambda$ production. This result is
qualitatively at variance with the conclusion that, e.g.,
the thrust and charged multiplicity distributions of
$\chi_{b,0}$ and $\chi_{b,2}$ two-gluon decays agreed well with
$\Upsilon$(1S)$\to ggg$ and that the thrust and charged multiplicity
distribution of $\chi_{b,1}\to q{\overline q}g$ agreed with
continuum $e^+e^-\to q{\overline q}$ events\cite{r:CLEO91}. This effect is
not reproduced in the current JETSET 7.4 Monte Carlo event
generator. For protons, which represent our
highest-statistics sample, our results are inconsistent
with a model where baryon production in gluon fragmentation
is only a function of the available center-of-mass energy.
Further detailed comparisons with models based on 
either string or independent parton fragmentation may help
clarify the production mechanisms.

\section{Acknowledgments}
We gratefully acknowledge the effort of the CESR staff
in providing us with excellent luminosity and running conditions.
D.~Cronin-Hennessy and A.~Ryd thank the A.P.~Sloan Foundation.
This work was supported by the National Science Foundation,
the U.S. Department of Energy, and
the Natural Sciences and Engineering Research Council of Canada.

\end{document}